\documentclass[fleqn,usenatbib]{mnras}

\usepackage{newtxtext,newtxmath}

\usepackage[T1]{fontenc}

\DeclareRobustCommand{\VAN}[3]{#2}
\let\VANthebibliography\thebibliography
\def\thebibliography{\DeclareRobustCommand{\VAN}[3]{##3}\VANthebibliography}

\usepackage{graphicx}	
\usepackage{amsmath}	
\usepackage{amsmath}
\usepackage{booktabs}
\usepackage[version=4]{mhchem}
\usepackage{xspace}
\usepackage{float}
\usepackage{makecell}
\usepackage{deluxetable}
\usepackage{starfont}
\usepackage{makecell}

\newcommand{\POSEIDON}{\texttt{POSEIDON}\xspace}

\title[Impacts of Stellar Contamination on Retrievals of TRAPPIST-1f]{Veiled in Starlight: Impacts of Stellar Contamination on Retrievals of TRAPPIST-1f's Atmospheric Composition}

\author[H. R. Rosenthal et al.]{Harper Rosenthal,$^{1,2}$\thanks{E-mail: crrosen@umich.edu}
Lisa Kaltenegger,$^{1}$
Ryan J. MacDonald$^{2,3}$,
Rebecca Payne$^{1,4}$,
and Elijah Mullens$^{1}$
\\
$^{1}$Department of Astronomy and Carl Sagan Institute, Cornell University, 122 Sciences Drive, Ithaca, NY 14853, USA \\
$^{2}$Department of Astronomy, University of Michigan, Ann Arbor, MI 48109, USA \\
$^{3}$School of Physics and Astronomy, University of St Andrews, North Haugh, St Andrews, KY16 9SS, UK \\
$^{4}$Department of Physics and Astronomy, Bates College, Lewiston, ME 04240
}

\date{Accepted XXX. Received YYY; in original form ZZZ}

\pubyear{\the\year{}}

\begin{document}
\label{firstpage}
\pagerange{\pageref{firstpage}--\pageref{lastpage}}
\maketitle

\begin{abstract}
The TRAPPIST-1 system offers seven terrestrial exoplanets with tight orbits and large radii ratios to the host star. If an atmosphere exists, transmission spectroscopy can be used to detect specific atmospheric features. Predictions of the atmospheric detectability of the TRAPPIST-1 planets prior to the launch of \textit{JWST} assumed pristine stellar surfaces. However, initial \textit{JWST} observations of the TRAPPIST-1 planets demonstrate that stellar contamination from unocculted active regions imparts significantly stronger spectral features than any planetary atmospheres. Here, we evaluate the atmospheric detectability of the habitable zone planet TRAPPIST-1f using atmospheric retrievals accounting for stellar contamination. We model a transmission spectrum given a CO$_2$-rich, habitable atmospheric model, and we include a ``worst case" stellar contamination spectrum. We then perform atmospheric retrievals on simulated \textit{JWST} observations with MIRI LRS (5-15 \micron) and NIRSpec PRISM (0.6-5.3 \micron), assuming accurate starspot spectral models. We find that NIRSpec observations alone achieve similar results as MIRI and NIRSpec together. We find $\sim$10 transits obtains strong evidence ($B>150$) for CO$_2$, and $\sim$50 transits finds weak evidence ($B>3$) for CH$_4$. We could not retrieve evidence of H$_2$O with up to 100 simulated transits with both instruments. Many challenges remain to accurately account for stellar contamination for ultra-cool M-dwarfs in atmospheric retrievals, and our results show that, while evidence for a CO$_2$-rich atmosphere around TRAPPIST-1f can be found with a short \textit{JWST} program, other prominent atmospheric signatures can only be disentangled from strong stellar features with more observation time than previous studies have indicated.

\end{abstract}

\begin{keywords}
exoplanets --- planets and satellites: individual: TRAPPIST-1f --- planets and satellites: atmospheres --- planets and satellites: terrestrial planets --- stars: activity --- techniques: spectroscopic
\end{keywords}



\section{Introduction}

\begin{figure*}
    \centering
    
    \includegraphics[width=\textwidth]{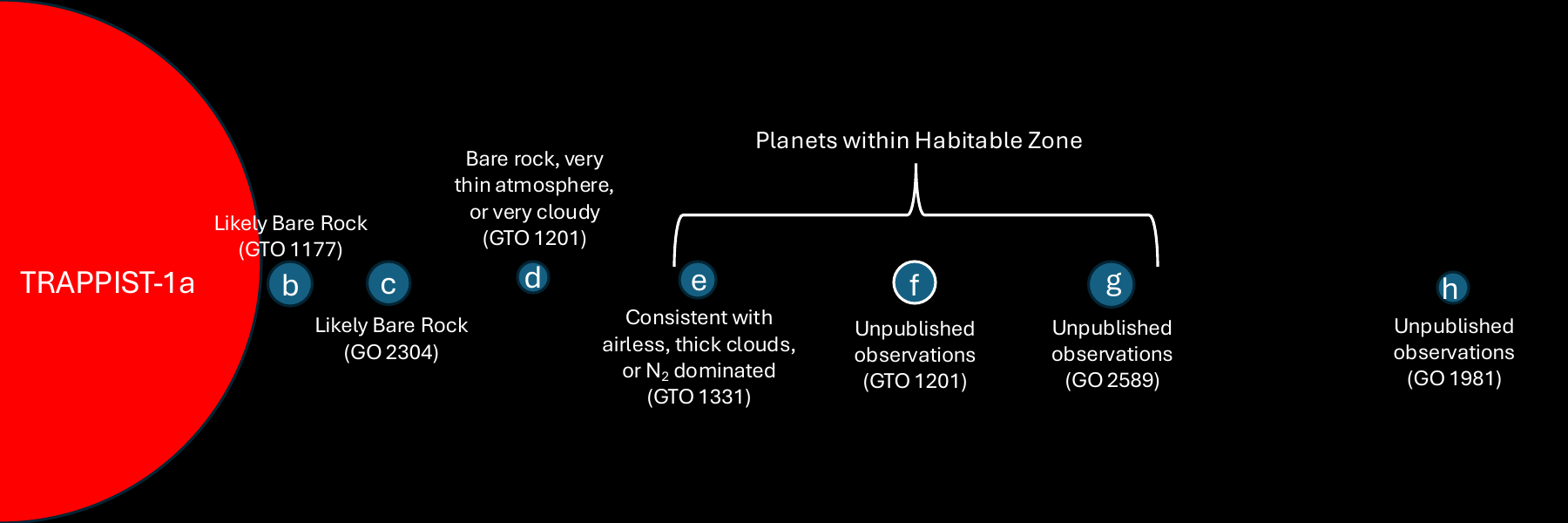}
    \caption{Illustration of the TRAPPIST-1 system and the most recent \textit{JWST} constraints on planetary atmospheres. Planet size and distance roughly to scale.}
    \label{fig:system_cartoon}
    \end{figure*}

Only six currently known rocky exoplanets can both be characterized by \textit{JWST} and are located within their star’s temperate Habitable Zone (HZ), the region in which it is possible to sustain liquid water on the surface of a rocky planet \citep{Cadieux2024,Kasting1993,Kopparapu2013,Ramirez2014}. The nearby TRAPPIST-1 system \citep[see Figure \ref{fig:system_cartoon}]{Gillon2016,Gillon2017} possesses seven rocky exoplanets orbiting an ultra-cool ($T_\mathrm{eff}$ $\simeq$ 2550 K) red dwarf star, including three that orbit in the HZ. TRAPPIST-1 is notable for being one of the best systems for observation of potentially habitable planets, due to a combination of proximity to Earth, frequent transits, and small radius of the host star \citep{Turbet2020}. Investigating the potential habitability of these potentially habitable planets, TRAPPIST-1e, TRAPPIST-1f, and TRAPPIST-1g \citep{Kopparapu2013}, has been at the forefront of modeling and observations \citep[e.g.][]{Turbet2018,Lin2021,Lim2023,Payne2024,Espinoza2025,Glidden2025}.

 The greatest challenge for the survival of an atmosphere in the TRAPPIST-1 system---and with it, any possibility of surface water that would be necessary for detectable surface-dwelling life--- is its star. The high stellar activity of M stars, and the extreme X/EUV radiation, stellar wind, and flaring of TRAPPIST-1 could have stripped any or all of its planets of any atmospheres \citep[e.g.][]{Ramirez2014,Wheatley2017,Airapetian2017, Dong2017,Dong2018,Garcia-Sage2017}. However, several theories demonstrate how exoplanets circling active M-stars like the TRAPPIST-1 exoplanets could have retained a habitable atmosphere despite the hostility of their host; for example, a sufficiently thick atmosphere on a rocky exoplanet could largely shield the surface from the high UV flux of an active M star \citep{OMalley-James2017,OMalley-James2019,Ramirez2014}, and H$_2$O could be protected from exacerbated photolysis if partially sequestered in the mantle \citep{Moore2020}, or if a larger water inventory exists or was delivered late \citep{Raymond2021,Ogihara2021}. Several teams \citep[i.e.][]{Turbet2018, Payne2024, Krissansen-Totton2024} modeled potential climates on the TRAPPIST-1 exoplanets and posited that several could sustain surface H$_2$O as long as the atmosphere contained a significant amount of CO$_2$ (on the order of $\geq1$1 bar). \citet{Payne2024} explored the system and identified several possibilities for habitable surface conditions for TRAPPIST-1e, TRAPPIST-1f, and TRAPPIST-1g and their corresponding CO$_2$ abundance and surface pressure. While the system has seen many observational missions with \textit{HST} and \textit{JWST}, further observations are needed to assess any possibility of temperate conditions and surface liquid water on the TRAPPIST-1 planets. 


\textit{JWST} provides the best opportunity for acquiring atmospheric data. Previous atmospheric retrieval studies have shown promising results for the system. Soon after the discovery of the TRAPPIST-1 planets, it was predicted that if the planets were to have Earth-like atmospheres, ozone could be detected with 30 transits observed with both NIRSpec and MIRI for TRAPPIST-1 c and d \citep{Barstow2016}. \cite{Meadows2023} indicate CH$_4$ and CO$_2$ would be detectable in as few as 10 transits for a modern biosphere on TRAPPIST-1 d and e. While \cite{Krissansen-Totton2018} finds 10 transits using NIRSpec PRISM can detect CO$_2$ and constrain CH$_4$ to rule out nonbiological production methods, \cite{Rotman2023} finds for TRAPPIST-1e that the number of transits needed to detect CH$_4$ is dependent on the CO$_2$ partial pressure, and therefore atmospheric temperature. \cite{Lin2021} finds 20 transits can additionally begin to constrain H$_2$O. \cite{Fauchez2019} simulated \textit{JWST} transmission spectroscopy of TRAPPIST-1 e, f, and g, finding that CO$_2$ was the most easily detectable molecule in a variety of concentrations and habitable scenarios, with a $3\sigma$ detection possible in less than 15 transits with NIRSpec, while no other gas was detectable in less than 100 transits. A similar conclusion is reached by \cite{Pidhorodetska2020}, who find that only CO$_2$ and its feature at $4.3\mu m$ would be reasonably detectable after including clouds in their simulated spectra of TRAPPIST-1e.
\cite{Barstow2016} found that present day Earth ozone levels would be detectable in 60 transits for TRAPPIST-1 b, 30 for c and d, with both NIRSpec and MIRI. However, these studies assumed a quiet stellar photosphere without contamination from active stellar regions such as starspots and faculae. 

The significant impact of stellar contamination motivates updated simulations to reassess the atmospheric detectability of the TRAPPIST-1 planets. Previous observations have been severely challenged by the star’s frequent and intense activity \citep[see e.g.][]{Lim2023,Howard2023,Rathcke2025,Radica2025}. \cite{Lim2023} observed different transmission spectra in their two visits of TRAPPIST-1 b\footnote{GO 2589} due to variable stellar contamination, showing strong evidence for an unocculted spot in the first and an unocculted facula in the second. Their observations could be described by stellar contamination alone, but better constraints on the stellar model are needed to completely rule out a thin or high mean-molecular-mass atmosphere. \cite{Radica2025} observed two transits of TRAPPIST-1 c$^1$, each displaying strong signatures of stellar contamination and consistent retrieved spot and faculae properties, noting that the sensitivity dropped by two orders of magnitude when stellar contamination was taken into account. The unknowns of the stellar contribution was more of a limiting factor than the precision of the transit depths. \cite{Piaulet-Ghorayeb2025}'s observations of TRAPPIST-1 d\footnote{GTO 1201}  were dominated by stellar contamination that varied between the two observed transits, and neither method used to account for the stellar surface were able to explain the stellar spectrum without caveats regarding model biases and mismatches with observational data. \cite{Espinoza2025} observed four transits of TRAPPIST-1e\footnote{GTO 1331}, noting stellar contamination over a wider wavelength range than previously reported and that current stellar models could not account for the spectral features arising from contamination. These observations highlight how accounting for this activity greatly changes what atmospheric constraints are reasonable, and is a critical component of any analysis for the TRAPPIST-1 planets. 

In this study, we focus on TRAPPIST-1f because of its potential to retain an atmosphere that allows habitable surface conditions \citep[e.g.][]{Payne2024} and due to its status as a priority \textit{JWST} target. It has also seen less exploration in both observational and theoretical studies, especially exploration that accounts for the stellar activity, but given prospects for habitability it deserves further attention. Previous observations of TRAPPIST-1f with \textit{HST} \citep{Zhang2018,DeWit2018} have yet to find find any planetary absorption features in the individual or combined spectra, but these observations were based on just 1 and 2 transits, respectively. Even then, \cite{Zhang2018} noted the presence of stellar contamination in the observation. \textit{JWST} program GTO 1201 also included 5 transits of observation of TRAPPIST-1f using NIRISS, but the results of these observations have yet to be published.

Here, we conduct an atmospheric retrieval analysis on simulated MIRI LRS and NIRSpec PRISM \textit{JWST} data for TRAPPIST-1f, including realistic stellar contamination informed from the first \textit{JWST} observations of the TRAPPIST-1 system.  In this study, we explore the impact of stellar contamination on atmospheric inferences for one of the most promising HZ planets accessible to \textit{JWST}. By using a habitable atmospheric model for TRAPPIST-1f, we investigate how quickly \textit{JWST} should be able to initially detect such an atmosphere, as well as determine how stellar contamination affects our ability to constrain potential biosignatures, such as the combination of O$_2$ or O$_3$ with CH$_4$ \citep[e.g.,][]{Kaltenegger2017,Kasting2014}. 

Our study is structured as follows. In Section~\ref{sec:methods}, we detail the atmospheric models used and the process of obtaining an atmospheric retrieval from the spectra of our model using \POSEIDON. In Section~\ref{sec:results}, we compare retrieval datasets and assess confidence in \textit{JWST}'s ability to constrain certain atmospheric components. Finally in Section~\ref{sec:discussion} we detail the significance of this analysis to future observations of TRAPPIST-1f.

\section{Methods}
\label{sec:methods}

\begin{table}
	\centering
    \tablewidth{\columnwidth}
    \tabletypesize{\scriptsize}
	\caption{Retrieval parameters, the reference values in the reduced pressure range, and the retrieval priors for that parameter. Parameters marked with a $^*$ are used in the stellar contamination retrieval only.}
	\label{tab:parameters}
	\begin{tabular}{lccr} 
        \textbf{Parameter} & \textbf{Description} & \textbf{Reference Value} & \textbf{Prior} \\
        \hline
        \hline
		$\mathrm{T}$ & Temperature & 163.075 K & $\mathcal{U}[100, 300]$ \\
        \hline
        $\mathrm{R_p}$ & Planet radius & 1.052 $\mathrm{R_{Earth}}$  & \makecell{$\mathcal{U}[0.894\,\mathrm{R_{Earth}},$ \\ $\,1.210\,\mathrm{R_{Earth}}]$} \\
        \hline
        ${\mathrm{\log \hspace{0.04cm} P_{surf}}}$ & Surface pressure & 0.762  & $\mathcal{U}[-6, 1]$ \\
        \hline
        $\log \hspace{0.04cm} \ce{H2O}$ & \makecell{Water vapor \\ mixing ratio}  & -5.697  & $\mathcal{U}[-12, 0]$ \\ 
        \hline
        $\log \hspace{0.04cm} \ce{O2}$ & \makecell{Oxygen \\ mixing ratio} & -1.456 & $\mathcal{U}[-12, 0]$ \\ 
        \hline
        $\log \hspace{0.04cm} \ce{O3}$ & \makecell{Ozone \\ mixing ratio} & -6.053 & $\mathcal{U}[-12, 0]$ \\ 
        \hline
        $\log \hspace{0.04cm} \ce{CH4}$ & \makecell{Methane \\ mixing ratio} & -4.664  & $\mathcal{U}[-12, 0]$ \\ 
        \hline
        $\log \hspace{0.04cm} \ce{N2O}$ & \makecell{Nitrous oxide \\ mixing ratio} & -6.457  & $\mathcal{U}[-12, 0]$ \\
        \hline
        $\log \hspace{0.04cm} \ce{N2}$ & \makecell{Nitrogen \\ mixing ratio} & -0.870 & $\mathcal{U}[-12, 0]$ \\
        \hline
        $\mathrm{F_{spot}}^{*}$ & \makecell{Sunspot \\ fractional coverage} & 0.23 & $\mathcal{U}[0.0, 0.5]$ \\
        \hline
        $\mathrm{F_{fac}}^{*}$ & \makecell{Faculae \\ fractional coverage} & 0.07 & $\mathcal{U}[0.0, 0.5]$ \\
        \hline
        $\mathrm{T_{spot}}^{*}$ & \makecell{Sunspot \\ surface temperature} & 2371 K & $\mathcal{U}[2300, 2631]$ \\
        \hline
        $\mathrm{T_{fac}}^{*}$ & \makecell{Faculae \\ surface temperature} & 2706 K & $\mathcal{U}[2391, 3013.2]$ \\
        \hline
        $\mathrm{T_{phot}}^{*}$ & \makecell{Photosphere \\ surface temperature} & 2571 K & $\mathcal{N}[2511, 40^2]$ \\
        \hline
        $\log \hspace{0.04cm} \mathrm{g_{spot}}^{*}$ & \makecell{Sunspot \\ surface gravity} & 4.55 & $\mathcal{U}[3.0, 5.4]$ \\
        \hline
        $\log \hspace{0.04cm} \mathrm{g_{fac}}^{*}$ & \makecell{Faculae \\ surface gravity} & 4.49 & $\mathcal{U}[3.0, 5.4]$ \\
        \hline
        $\log \hspace{0.04cm} \mathrm{g_{phot}}^{*}$ & \makecell{Photosphere \\ surface gravity} & 5.24 & $\mathcal{N}[5.2396, 0.1^2]$ \\
	\end{tabular}
\end{table}

\begin{figure*}
    \centering
    \includegraphics[width=\textwidth]{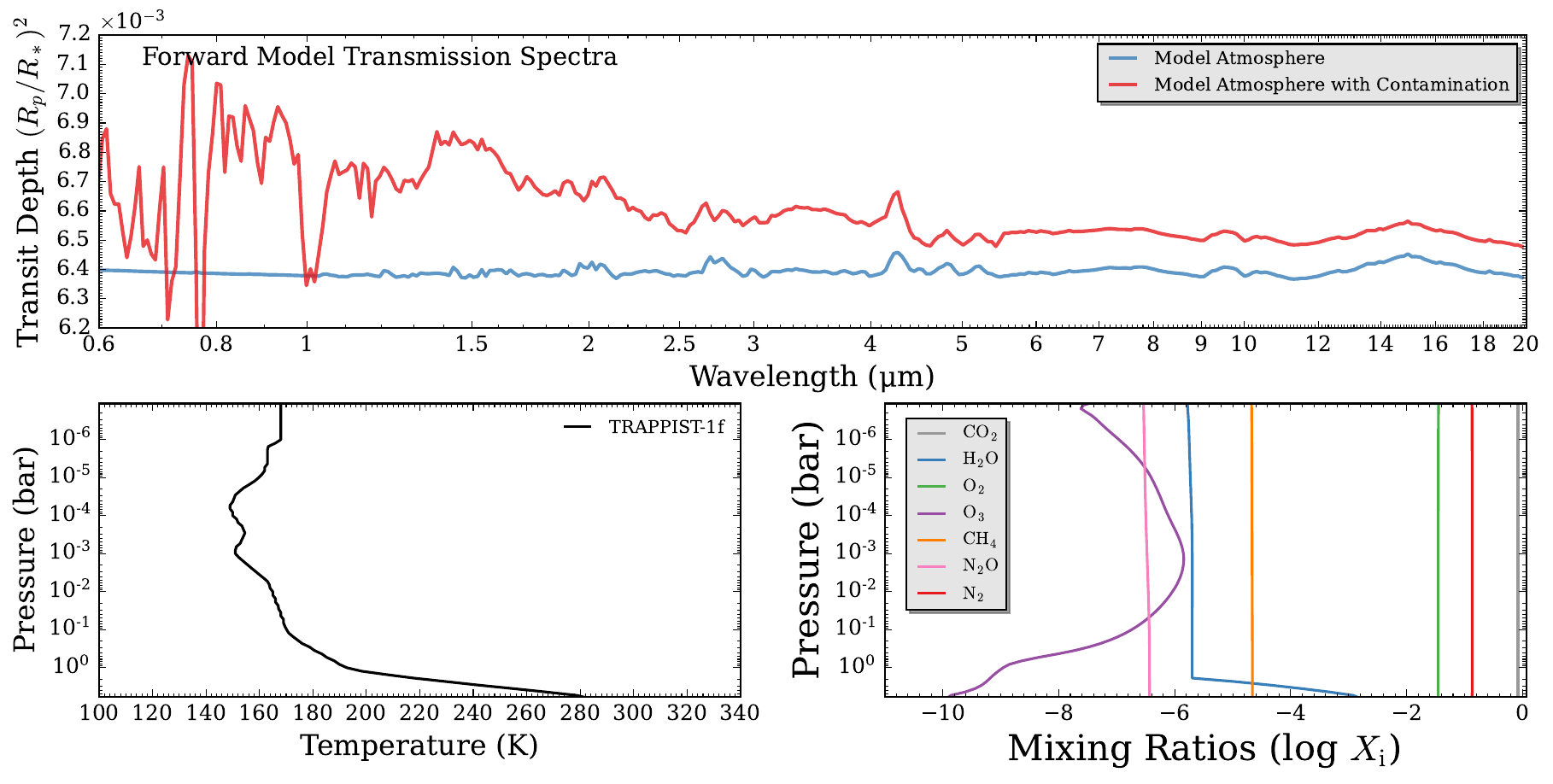}
    \caption{(Top) TRAPPIST-1f transmission spectra for the model atmosphere alone (blue) and the model when stellar contamination is included (red) \citep{Lim2023}. (Bottom left) Atmospheric 5-bar surface pressure model for TRAPPIST-1f \citep{Payne2024}. (Bottom right) Atmospheric chemistry model for TRAPPIST-1f in the 5-bar surface pressure model \citep{Payne2024}.}
    \label{fig:model}
\end{figure*}

In this section, we discuss the structure and makeup of the atmospheric model we use for TRAPPIST-1f, and explain how we use the \POSEIDON package to generate contaminated and non-contaminated transmission spectra. We then discuss the creation of simulated \textit{JWST} observational data from these model spectra. Finally, we discuss the process of atmospheric retrievals to analyze the simulated data. 

\subsection{Forward Model Transmission Spectra}

\cite{Payne2024} explored habitable scenarios for TRAPPIST-1f and generated several atmospheric models that could lead to a habitable surface temperature. We use the lowest pressure atmosphere model identified in \cite{Payne2024} that allows for temperate surface conditions for TRAPPIST-1f (see bottom panels of Figure \ref{fig:model}), that being an Earth-like atmospheric model with an added $\sim$5 bars of pressure from CO$_2$. We chose this model to explore a ``best-case scenario" where enough CO$_2$ is present that the planet can sustain a habitable temperature with minimal obscuring of other atmospheric features. We use planetary and stellar parameters from \cite{Delrez2018} and \cite{Grimm2018}: $R=1.052~\mathrm{R_{Earth}}$, $M=\mathrm{0.934~M_{Earth}}$, and a density of $0.816$ Earth densities. 

We create forward model atmospheric spectra from this atmospheric model using the open-source \POSEIDON \footnote{Repository: https://github.com/MartianColonist/POSEIDON\\Documentation: https://poseidon-retrievals.readthedocs.io} package \citep{MacDonald2017,MacDonald2022,MacDonald2023}, a Python package that couples a parametric planetary atmosphere and radiative transfer model to a Bayesian parameter estimation and model comparison suite built on \texttt{MultiNest}, adapted to rapidly retrieve atmospheric properties from terrestrial exoplanet spectra \citep{Kaltenegger2020,Lin2021}. 

\begin{figure*}
    \centering
    \includegraphics[width=\textwidth]{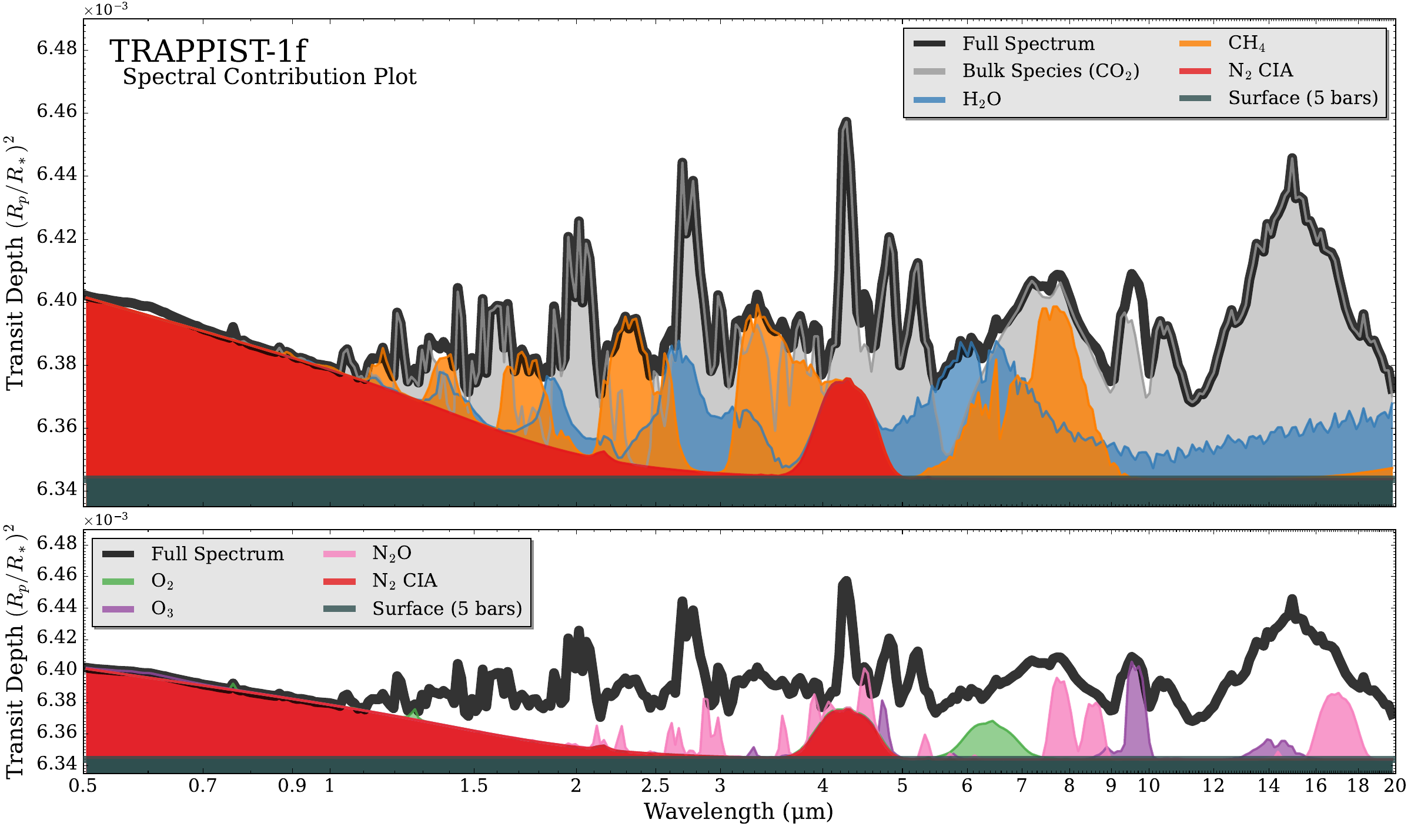}
    \caption{The transmission spectra for the TRAPPIST-1f Earth-like 5-bar atmosphere model, without stellar contamination. Individual opacity contributions of major spectrally active molecules are shown (colored curves) relative to the spectral continuum due to Rayleigh scattering and refraction. Collision-induced absorption (CIA) pairs featuring O$_2$ (O$_2$-O$_2$ and O$_2$-N$_2$) are depicted alongside the O$_2$ contribution. Molecular contributions are displayed in two panels for clarity.}
    \label{fig:contribution}
\end{figure*}

In our forward model, we choose to include in our forward model several spectrally prominent molecules at the wavelength ranges of NIRSpec PRISM and MIRI LRS that are expected in an Earth-like habitable atmosphere : N$_2$, H$_2$O, CO$_2$, N$_2$O, O$_2$, O$_3$, and CH$_4$ \citep{Lin2021}, specifying CO$_2$ as the bulk gas in all models. The molecular cross sections used by \POSEIDON for temperate planets \citep{Kaltenegger2020} are derived from HITRAN 2020 line lists \citep{Gordon2022}, with the addition of optical O$_3$ absorption \citep{Serdyuchenko2014}, the latest HITRAN collision-induced absorption (CIA) data \citep{Karman2019}, and additional line lists for CO$_2$, CH$_4$, \citep{Yurchenko2020}, H$_2$O \citep{Polyansky2018}, and N$_2$O \citep{Hargreaves2019}. \POSEIDON interpolates the model mixing ratio profiles and atmospheric temperature structure onto a 100-layer vertical grid spaced uniformly in log-pressure from -9.0 to 2.0 in log-pressure, with a hard surface at 0.762 in log-pressure \citep{Lin2021,Payne2024}. We generate model transmission spectra with a resolution of R=10000 from 0.5 to 20 $\mu$m for all modeled scenarios, as shown in the top panel of Figure \ref{fig:model} (blue). Figure \ref{fig:contribution} depicts the key spectral contributions of each molecule towards our model's transmission spectra.

We also consider the impact of stellar contamination on the transmission spectra. The stellar contamination module available in \POSEIDON has been used extensively to interpret transmission spectra of M-dwarf planets (Recent examples found in \cite{Bennett2025,Moran2023,May2023}). We use a two-heterogeneity model to represent this contamination: one starspot colder than the photosphere, and one hotter faculae. We supply the model with the fractional coverage, temperature, and surface gravity of the starspots and facula using the median retrieved parameters from \cite{Lim2023}, found in Table \ref{tab:parameters}. Typically, unocculted spots create positive features that mimic absorption or scattering in the spectra, while unocculted faculae create negative features that can hide real spectral features \citep{Rackham2018}. \POSEIDON's starspot model mimics water absorption features with similar amplitude to actual atmospheric features, and, for simplicity, we assume the model holds for each transit. Then, by using the \texttt{PyMSG} package \citep{Townsend2023}, we interpolate the PHOENIX stellar models \citep{Husser2013} to calculate the adjustment of the stellar contamination to the transmission spectra \citep{Lim2023}. Due to the limits of the grid, the stellar contamination can only be modeled up to 5.5 µm, after which the atmosphere-only spectra is scaled by a constant value due to temperature differences between the spot/faculae and photosphere. We choose to use PHOENIX over a stellar model that extends to higher wavelengths such as SPHINX \citep{Iyer2023} for a few key reasons. While SPHINX extends past 5.5 µm and allows for a lower temperature prior bound, the lower spectral resolution (R$\sim250$) and limited log \textit{g} values (4.0-5.5 dex) of the models would still result in limitations on the accuracy of our stellar models. Additionally, creating a forward model of our contaminated spectra using SPHINX instead of POSEIDON resulted in a spectra with negligible differences past 5.5 µm (within the margin of error of the \texttt{PandExo} simulated data points, see Section \ref{sec:SUBSEC_simulated_obs}). Given both models have their limitations, and PHOENIX has been well-tested for this use case, we choose to use PHOENIX models only for this study, and comparisons with results using SPHINX are grounds for future analysis. For more discussion of comparison of theoretical stellar spectra to observations and the differences between PHOENIX and SPHINX, refer to Appendix D of \cite{Lim2023}. Lastly, for the retrievals we assume an error of 40 K for photosphere temperature and 0.1 for log of surface gravity. We adopt this conservative value to explore a wide variety of photosphere temperatures, while still staying inside \texttt{PyMSG}'s calculated stellar grid bounds. 

The top panel of Figure \ref{fig:model} compares the transmission spectra with (red) and without (blue) stellar contamination, assuming that stellar contamination remains the same for each transit. Note that this is not the case in real observations \citep[see e.g.][]{Lim2023}, and we make this assumption in order to determine how constraints on TRAPPIST-1f’s atmosphere change with the introduction of stellar contamination, in the limit where it is present but not strongly time-variable. While this is a major assumption, it has observational precedent in that \cite{Radica2025} observed consistent stellar properties between visits of TRAPPIST-1 c a year apart, although these properties were largely unconstrained. Additionally, current observations of the system are still limited by incomplete knowledge of TRAPPIST-1's behavior; for example, \cite{Piaulet-Ghorayeb2025} finds the stellar models to be a limiting factor and \cite{Espinoza2025} observes stellar contamination over a wider wavelength range than current models are able to completely explain. Constant stellar properties also make retrievals with many transits of simulated data computationally feasible. Modeling stellar contamination as identical for each transit therefore serves as a baseline for understanding our observational limits for TRAPPIST-1 f, and more work is required to understand real-world observations. For further discussion, see Section \ref{sec:SUBSEC_sensitivity}.

\subsection{Simulated Observational Data} \label{sec:SUBSEC_simulated_obs}
We use the \texttt{PandExo} Python package to create synthetic \textit{JWST} observational data for each modeled scenario. \texttt{PandExo} takes the modeled spectra and returns a transit depth error and bin size of each pixel, adding noise and error to our ground truth model \citep{Batalha2017}. We simulate observations with the NIRSpec PRISM (0.6-5.3 µm) and MIRI LRS (5-15 µm) instruments, as the spectral features present at these wavelengths are key for investigating habitability \citep{Kaltenegger2009}, with saturation at 80\% full well. We assume observation for the entire transit, with a transit time of 63.14 minutes \citep{Delrez2018}, and a total observation time of 2 times the transit time plus the transit observation \citep{Lin2021}. For NIRSpec PRISM, we had 2 groups per integration with a mean error of about 110.29 ppm, and for MIRI LRS, we had 139 groups per integration, with a mean error of about 301.48 ppm. Additionally, we re-binned the MIRI LRS data produced to 0.25 µm-wide bins, consistent with previous work using atmospheric retrievals on MIRI data \citep{Grant2023}. 

We simulated our observational data without Gaussian scatter. Especially for our retrievals with fewer transits, this may cause our results to be overly optimistic about the constraints of our posteriors. We discuss the possible changes in retrieved parameters that could result from different Gaussian scatter draws in Section \ref{sec:SUBSEC_sensitivity}. Our results for these few-transit simulated observations should be taken as a "best-case" baseline, with the possibility of shifts in the precision and constraints of our retrievals.

We simulate a 1 transit \textit{JWST} observation of the planet with each instrument, then scale down the error bars by a factor of $1/\sqrt{N}$ to create the datasets simulating $N$ transits with that instrument. 

\subsection{Atmospheric Retrieval}

We use \POSEIDON’s atmospheric retrieval code to extract atmospheric properties from our simulated model \textit{JWST} observation. As described in \cite{Kaltenegger2020}, \POSEIDON has been adapted from work on gas giants to model and retrieve transmission spectra of terrestrial planets, adding retrieval parameters for surface pressure and including molecular cross sections for temperate planets \citep{Kaltenegger2020, Lin2021}. \POSEIDON couples a parametric planetary atmosphere and radiative transfer model to a Bayesian parameter estimation and model comparison suite built on \texttt{MultiNest} \citep{Feroz2008, Feroz2009, Feroz2019} via the \texttt{PyMultiNest} package \citep{Buchner2014}. \texttt{MultiNest} ran each retrieval using 2000 live points.

To run the retrieval, we use uniform priors for isothermal temperature (100-300 K), planetary radius (0.85 to 1.15 times TRAPPIST-1f’s), and the log of the surface pressure (-6.0 to 1.0). We cannot go lower than 100~K for our temperature prior as that is the lowest temperature available in the opacity grids, as well as being extremely cold for a realistic TRAPPIST-1f. we first use the centered-log-ratio (CLR) probability distributions for the priors, a parametrization that suits a Bayesian retrieval in which the main atmospheric molecular contribution is not exactly known since the results are independent of the order in which each molecule is sampled \citep{Benneke2012}. 

The CLR transformation of the \textit{i}th gas in \POSEIDON is given by $\xi_i=\ln{\frac{X_i}{g(\textbf{x})}}$ where $g(\textbf{x})=(\Pi^n_{j=0}X_j)^{1/n}$ \citep{Lustig-Yaeger2023}. For n=6 gases and defining a minimum mixing ratio at $X_{min}=-12$, the transformed variables sample over a uniform prior of $\approx \mathcal{U}[-22.148, 23.684]$. The upper limit of this prior corresponds to the \textit{i}th gas dominating the atmosphere where all other $X_j=X_{min}$, while the lower limit corresponds to $X_i=X_{min}$ with the other gases filling the remainder of the atmosphere. This treatment of the CLR transformation differs slightly from that described in \cite{Piette2022} and \cite{Welbanks2021}, which use a lower bound for all but one of the \textit{i} gases. \cite{Bello-Arufe2025} shows, in its use of the Aurora retrieval code in comparison to \POSEIDON, that the Welbanks treatment creates artificial lower limits on the mixing ratios, while the \POSEIDON CLR priors allow the posterior to remain flat all the way down to the lower bound, as we expect for non-detections.

For the data including stellar contamination, we include additional stellar parameters for the surface gravity at the photosphere and the temperature, surface gravity, and fractional coverage of the starspots and faculae. \POSEIDON's default values for the spacing of the stellar tempreature and log surface gravity grid is not precise enough for retrievals exploring stellar contamination. Therefore, we set the spacing of the stellar temperature grid to 5 and the log surface gravity grid to 0.02. All priors for all parameters used in our retrievals are shown in Table \ref{tab:parameters}.

For each model (uncontaminated and contaminated) we ran observations using 5, 10, 15, 25, 50, and 100 transits per instrument with both MIRI LRS and NIRSpec PRISM. For context, \textit{JWST} Cycle 3’s GO-6456 proposal to observe 15 transits of TRAPPIST-1e with NIRSpec PRISM was approved to take 128.8 hours in total \citep{Allen2024}, and 130+ hours is considered a "large" proposal with more intensive vetting. In this work, we explore using more transits to investigate how much observing time would be realistically needed in order to confirm the presence of certain molecules in a habitable scenario, as well as to investigate if the influence of stellar contamination can be diminished with more transits.

For each simulated observation, we also conducted a retrieval omitting the simulated MIRI LRS data, as many of the significant spectral features we anticipate detecting are in the NIRSPEC wavelength and the MIRI LRS data has a significantly larger margin of error. However, the wavelength range MIRI operates in is safe from stellar contamination, making its significnace to our results more debatable. Determining if observing with the MIRI instrument is redundant to our results will inform future observations since requiring MIRI data would greatly increase the number of transits needed.  

Additionally, for each combination of model, number of transits, and instrument variation, we conduct two additional retrievals: one excluding CH$_4$ and one excluding H$_2$O from the retrieval model. We then conduct a Bayesian model comparison with the full-model retrieval to find the detection significances. We chose to examine the evidence for CH$_4$ and H$_2$O specifically because, other than CO$_2$, they are the most spectrally prominent molecules in our forward model spectra, as well as being key for understanding habitability. If our retrievals fail to find significant evidence of CH$_4$ and H$_2$O in our data, it is reasonable to assume the remaining, much less spectrally obvious molecules will be increasingly difficult to detect. We also conduct retrievals removing CO$_2$ for the 5+5 and 10+10 transit cases to establish a baseline of detection for our most prominent molecule, indicating N$_2$ as our bulk gas in our retrieval model. Table \ref{tab:results} visually depicts the retrievals conducted with all  combinations of model, observational parameters, instruments, and excluded molecules.

\section{Results}
\label{sec:results}

To explore how stellar activity influences detectable atmospheric features of rocky Earth-like planets, we focus on TRAPPIST-1f and explore the impacts of stellar contamination on retrievals of the atmospheric composition for a temperate Earth-like atmosphere model. Fig. \ref{fig:model} shows our atmospheric model with stellar contamination (red) and the atmospheric-only model for TRAPPIST-1f (blue). 

\begin{figure*}
    \centering
    \includegraphics[width=\textwidth]{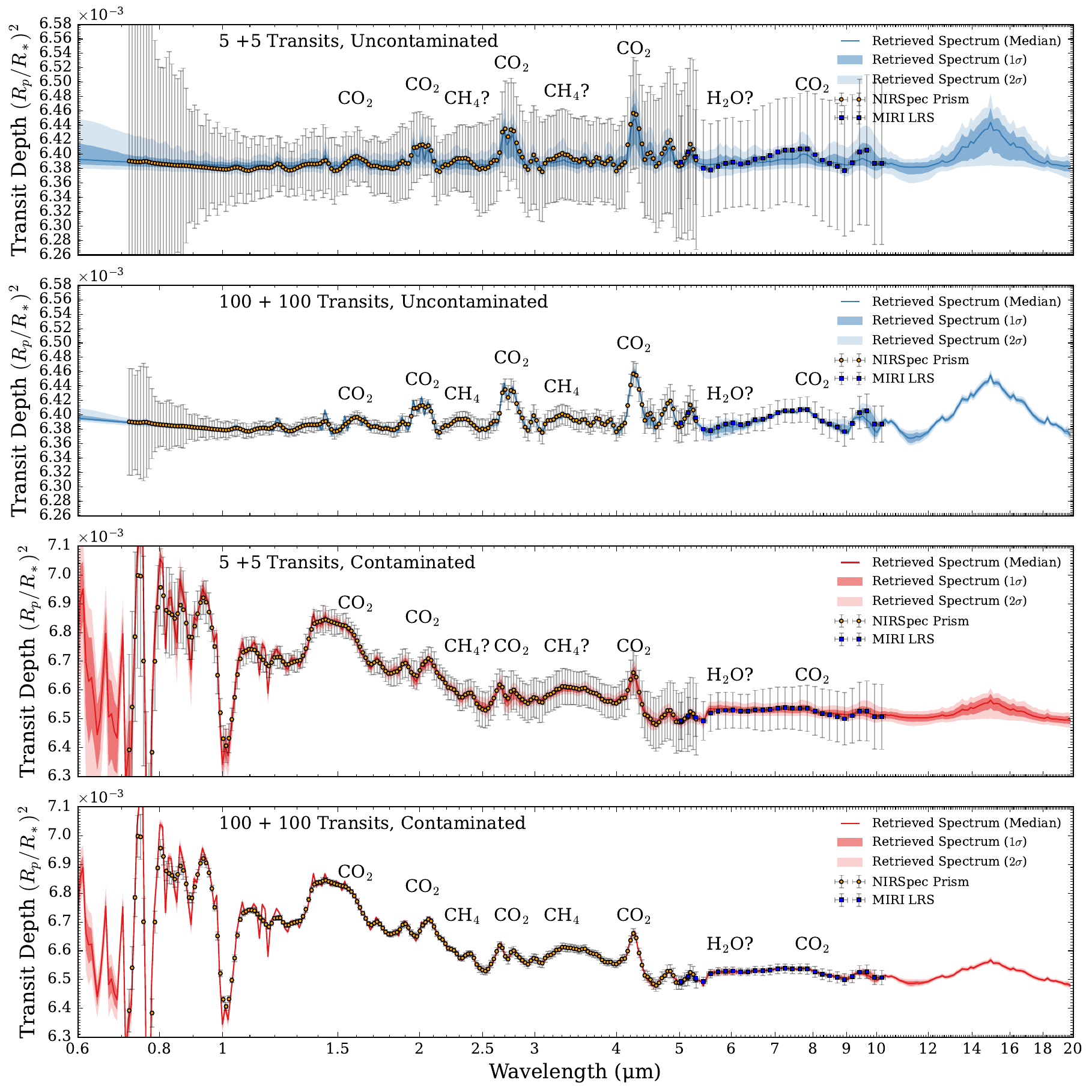}
    \caption{Retrieval spectra for the 5 + 5 and 100 + 100 transit retrievals of TRAPPIST-1f using the atmosphere-only model (blue) and the model including stellar contamination (red). The unscattered, \texttt{PandExo} simulated \textit{JWST} data is shown (0.6-5.3 µm NIRSpec PRISM data, 5-15 µm MIRI LRS data). Text labels indicate the locations of the most visually apparent spectral features. Features labeled with a question mark indicate the retrieval was unable to find sufficient evidence for that feature.}
    \label{fig:retrieved_spectra}
\end{figure*}

\begin{figure*}
    \centering
    \includegraphics[width=\textwidth]{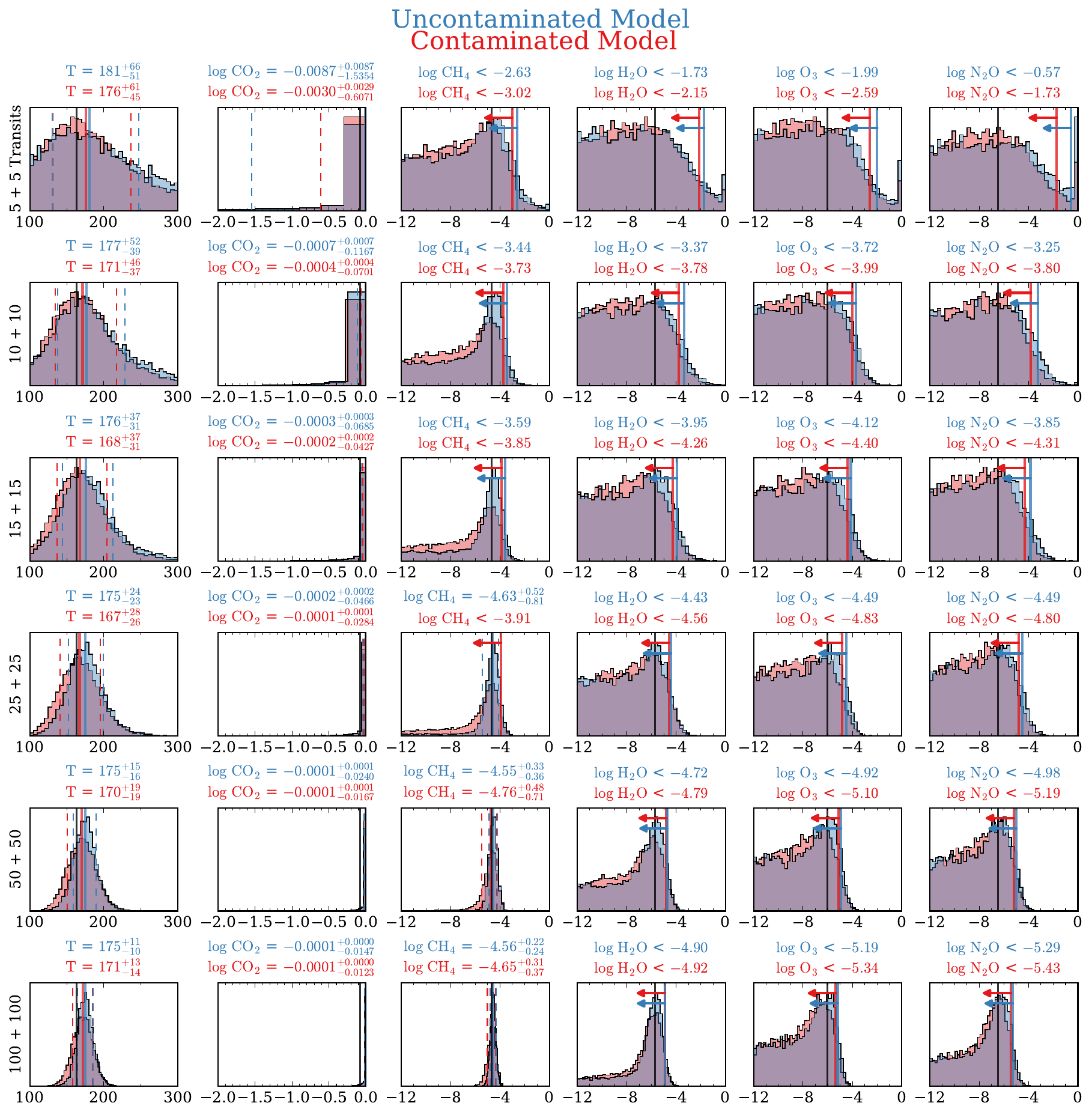}
    \caption{Retrieved properties of the models. Each panel shows the results from the atmosphere-only model (blue) and model including stellar contamination (red). The black vertical line represents the average value for that parameter in the 100 millibars to 0.1 millibars range. Vertical dotted lines represent the 1$\sigma$ uncertainty for constrained parameters, and arrows indicate the 95\% upper limit for unconstrained parameters. The histograms for the CO$_2$ mixing ratio show additional decimal places for the posterior distributions to avoid rounding to zero.}
    \label{fig:retrieved_histograms}
\end{figure*}

The forward model used was not isothermal and assumed chemical equilibrium (refer to Figure \ref{fig:model}), while the retrieval used isothermal and iso-composition profiles; therefore, in order to compare our retrieved isochemical abundances to the forward model, we compare the retrieved values to the average of the value in the height range of the atmosphere from 100 millibars to 0.1 millibars (refer to Figure \ref{tab:parameters} for the values used). For the parameters that do not vary with transit depth, namely the stellar parameters, planetary radius, and surface pressure, the values supplied to the initial model are our reference values. These reference values are listed in Table \ref{tab:parameters}. We choose to use isochemical and isothermal profiles for our retrievals despite our forward model not being isothermal/isochemical to better simulate what is done in practice for these observations, that being a realistic atmosphere being retrieved by an isochemical profile to allow the data to guide best-fit parameters. This may result in improved precision due to model assumptions, but makes our results more applicable to real-world observations.
 
Our retrieved spectra for the 5 NIRSpec + 5 MIRI and 100 + 100 transit simulations on the atmosphere-only and contaminated models are shown in Figure \ref{fig:retrieved_spectra}. Even with the dramatic decrease in error at the 100 transit upper limit, any atmospheric spectral features $<1.5 \mu m$ are obscured, and previously prominent signals of CO$_2$ and CH$_4$ between 1.5 and 3 $\mu m$ are distorted. However, based on the retrieved posterior distributions (Figure \ref{fig:retrieved_histograms}), 5 + 5 transits with \textit{JWST} is able to constrain CO$_2$ abundance, for both the contaminated and uncontaminated retrievals. We also find that the retrievals were able to constrain the stellar contamination parameters accurately (see Figure \ref{fig:stellar_param_retrieval} in the Appendix); however, these results rely on the assumption of the star remaining unchanged between visits. Our ability to retrieve stellar contamination parameters when the star is changing is addressed in Section \ref{sec:SUBSEC_sensitivity}.

\subsection{Bayesian Comparison Tests}

For the Bayesian comparison tests, comparing retrieval results with a retrieval with a molecule removed from the retrieval model, we classify our results with the following terms: ``no evidence" (B$_{ref}$ $<$ 3), ``weak evidence" (3 $<$ B$_{ref}$ $<$ 12), ``moderate evidence" (12 $<$ B$_{ref}$ $<$ 150), ``strong evidence" (150 $<$ B$_{ref}$ $<$ 600), and ``detection" (B $>$ 600) \citep{Schmidt2025}. Further discussion on the proper use of Bayesian evidence in retrievals can be found in \cite{Thorngren2025}. All Bayesian comparison tests and their exact Bayes factors can be found in the Appendix, Table \ref{tab:results}.

Looking first at the retrievals for the atmosphere-only, uncontaminated model and dataset, we find weak evidence for CO$_2$ at 5+5 transits ($B_{ref}=5.72$), strong evidence at 10+10 transits ($B_{ref}=289$), and a detection of CO$_2$ (relative to an N$_2$ dominated atmosphere) at 15+15 transits ($B_{ref}=2.38*10^4$). We find that at 5 + 5, 10 + 10, and 15 + 15 observed transits, there is no Bayesian evidence for the presence of CH$_4$ or H$_2$O. With 25 + 25 transits, weak evidence for CH$_4$ at $B_{ref} = 4.23$ is retrieved, increasing to $B_{ref} = 120$ at 50 transits (``moderate evidence") and a confident detection, $B_{ref} = 1.09 * 10^5$, at 100 transits (``detection"). Very notably, for retrievals where MIRI LRS data is not included, the variation in significances is extremely minor, implying that only NIRSpec PRISM data is necessary to constrain CH$_4$ in the uncontaminated scenario. For H$_2$O, given the low abundance in our atmospheric model, no retrieval, up to 100 + 100 transits of simulated observation, resulted in Bayesian evidence $>3$, reaching a maximum of only $B_{ref}=1.82$. We will note that the retrievals without MIRI LRS data had a notably lower Bayes factor (maximum of $B_{ref}=0.639$), so it does appear that the instrument has a significant role in constraining water abundance. We also attempt a Bayes test with retrievals excluding O$_2$ and O$_3$ with the 100+100 transit simulated data, finding no Bayesian evidence for either molecule. We can assume that given this best-case scenario, O$_2$ ($B_{ref}=1.03$) and O$_3$ ($B_{ref}=0.697$) are not detectable in any of our scenarios.

For the retrievals with stellar contamination included, we find similar results for CO$_2$, with weak evidence at 5+5 transits, strong evidence at 10+10 transits, and a detection at 15+15 transits ($B_{ref}=6.26$, $B_{ref}=197$, and $B_{ref}=2.38*10^4$, respectively). However, the retrieval is not able to find even weak evidence for CH$_4$ up to the 25 + 25 transit retrieval. For the 50 + 50 transit retrieval Bayesian comparison finds weak evidence ($B_{ref}=5.36$), and 100 transits finds strong evidence for CH$_4$ ($B_{ref}=177$). As with the uncontaminated retrievals, removing the MIRI LRS data has negligible effects on CH$_4$. With 100 + 100 transits, the contaminated retrievals are unable to find evidence for H$_2$O, reaching a maximum Bayes factor of $B_{ref} = 0.913$. Retrievals removing MIRI LRS once again notably lower the Bayes factor. Similarly to the above, a test for O$_2$ ($B_{ref}=0.642$) and O$_3$ ($B_{ref}=0.568$) at the 100 + 100 transit level yields no evidence for either.

\subsection{Sensitivity Tests} \label{sec:SUBSEC_sensitivity}

Our results rely on the simplifications that the star remains unchanging between visits and the choice to omit Gaussian scatter from our data. Here, we test our results for the impacts of varying our stellar contamination parameters or adding Gaussian scatter.

We conducted two tests of the validity of our results under varying stellar conditions. First, we randomized the stellar contamination input parameters for the starspot and faculae uniformly within the priors of \cite{Lim2023} (see Table \ref{tab:parameters}) to create five instances of randomized stellar parameters, to represent five observations of a changing star. We then perform retrievals on simulated observational data using these stellar parameters. For each set of randomized parameters, we simulate data from a single transit observation, and run one retrieval using NIRSpec alone and one using MIRI alone. We find that for our five stellar models, the posterior distributions of the atmospheric features are completely unconstrained for both NIRSpec and MIRI. For NIRSpec, the stellar parameters themselves tended to be well retrieved, opening the possibility for future work to explore using the constraints made on the stellar parameters with one transit to model and remove the stellar contamination from the spectra. However, given that stellar contamination in our model does not affect the spectrum beyond 5.5$\mu$m, the MIRI retrievals were not able to constrain the stellar parameters for the starspot and faculae.

To expand this exploration to explore a changing star during multiple transits of observation is currently not computationally feasible. With \POSEIDON, we would need to allow the six stellar parameters for the starspots and faculae to act as free parameters for each individual transit, which would make the retrievals unfeasible in \texttt{MultiNest}. For example, 5 transits of data with a single instrument would require 31 free parameters for the star alone, and as this number grows with more transits or instruments, it becomes impossible for \texttt{MultiNest} to reasonably accomplish. However, future retrieval codes may be able to address this.

Additionally, in order to highlight that the results presented in this paper are independent of the specific "draw" of stellar parameters observed by \cite{Lim2023}, we run a 15+15 transit retrieval with the same setup as our primary 15+15 contaminated retrieval, but using the same five instances of randomized stellar parameters that we used above. We see that there are no significant changes in the posterior distributions of radius, temperature, surface pressure, or atmospheric mixing ratios between our original 15+15 transit retrieval and the retrievals with randomized stellar parameters. Therefore, our atmospheric results, within the assumption of unchanging stellar parameters, is likely to be independent of the specific stellar parameters chosen.

We examined the effect of including Gaussian scatter in our data for our 5+5 transit case, where scatter should have the greatest impact on our results, with both the noncontaminated and the contaminated model. For both, we created five instances of Gaussian scatter and ran our retrieval on those datasets with the same setup as for the unscattered data. For both cases, we find that the retrieved posteriors for the surface pressure, as well as the mixing ratios of H$_2$O, CH$_4$, O$_3$, and N$_2$O, were sensitive to Gaussian scatter. For the contaminated case, the retrieved posteriors for the fractional coverage of the faculae also varied in different Gaussian scatter instances. All other parameters remained consistent with their nonscattered posterior distributions. Figures \ref{fig:appendix_uncontam_gaussian_scatter} and \ref{fig:appendix_contam_gaussian_scatter} in the Appendix show the posterior distributions for the affected parameters. Compared to the unscattered posterior distribution in the leftmost column, we see that the runs without Gaussian scatter still represent the typical result from retrievals with different noise instances. However, we do see that a noise draw can result in an erroneous result, best shown by O$_3$ in the third column of the uncontaminated case (Fig. \ref{fig:appendix_uncontam_gaussian_scatter}. See also H$_2$O in the third column, CH$_4$ in the sixth column of Fig. \ref{fig:appendix_contam_gaussian_scatter}). Given that our results show O$_3$ and H$_2$O are not detectable even in the best-case, non-scattered retrievals at these low transit observations, it is unlikely that our detection significance results would change significantly with the addition of Gaussian scatter. However, with the assumption of unscattered data, our results should still be interpreted as a "best-case" scenario for these simulated observations.

\section{Conclusions}
\label{sec:discussion}

Atmospheric observations of the TRAPPIST-1 planets with \textit{JWST} are currently probing our understanding of the environment of exoplanets in the Habitable Zone, giving us new insights into terrestrial planets and their evolution. However, accounting for stellar activity is critical in our ability to retrieve atmospheric properties, including potential biosignatures. 

In this work we use a modeled CO$_2$-rich, potentially habitable atmosphere for TRAPPIST-1f and create model transmission spectra, one scenario that is atmosphere-only and one scenario where we model the effect of stellar contamination. With the \texttt{PandExo} package, we simulate \textit{JWST} observations using NIRSpec PRISM and MIRI LRS from 5 + 5 transits to 100 + 100 transits. We then conduct atmospheric retrievals on the simulated observational data using \POSEIDON, including stellar parameters in our retrieval for the contaminated forward model spectra, to determine how many transits are required for \textit{JWST} to identify the key features of the forward model spectra. 

Several teams have investigated the contribution of stellar contamination from unocculted star spots and/or faculae \citep{Lim2023,Rackham2018,Rackham2017,McCullough2014, Wakeford2019,Garcia2022,Zhang2018,Rackham2024}. We adopt the model proposed by \citep{Lim2023} in our analysis. This contamination model assumes that the visible stellar disk has two unocculted regions, one with a temperature lower than the quiet photosphere (spot), and one with a higher temperature (facula). Results for different stellar models could influence the retrieved values, however, we have adopted the best model results from the observations from \citep{Lim2023} for TRAPPIST-1 for consistency with the data to investigate its effect on the possibility of retrieving atmospheric properties, including potential biosignatures. 

Note that our results may be impacted by some key assumptions in our modeled data. Due to computational limits, we assumed an unchanging star between transits. While our results do not seem to depend on the specific values of the stellar parameters chosen for our model, our ability to model the retrievals of TRAPPIST-1f for models of a star that changes between transits will be a future implementation. Additionally, we did not apply Gaussian scatter to our data to ensure our results were not biased towards a specific scattering of our data. However, we acknowledge that this may result in overly optimistic retrieval precision and constraints, especially at the lower number of transits where our error is higher (e.g., the 5+5 cases). Our results especially at those lower number of transits should be treated as an optimistic, best-case scenario.

We find that:
\begin{itemize}
    \item Stellar contamination significantly obscures any spectral features $<1.5$ \micron.
    \item Weak evidence for CO$_2$ ($B>3$) is retrieved at 5+5 transits in both contaminated and noncontaminated cases, with strong evidence at 10+10 ($B>150$) and a confident "detection" (compared to a retrieval model without CO$_2$ included) at 15+15 transits ($B>600$).
    \item Evidence for CH$_4$ ($B>3$) is retrieved at 25 + 25 transits for the uncontaminated model and retrieval, 50 + 50 transits in the contaminated model and retrieval.
    \item Evidence for H$_2$O is not found in up to 100 + 100 transits for any retrieval. The abundances of the remaining species in our model are even less constrained.
    \item Performing the atmospheric retrievals on the NIRSpec PRISM data alone, ignoring MIRI, had no effect on the above results, indicating that similar constraints on the atmospheric composition can be made with half of the observing time.
\end{itemize}

In comparison to the simulations that do not take stellar contamination into account, we show that once stellar contamination is accounted for, the observational predictions may vary significantly. For CH$_4$, analysis without considering stellar contamination in the forward model or retrieval results in a prediction of finding initial evidence in roughly half the observing time, highlighting the necessity of accounting for contamination when making predictions for observational plans or data analysis.

Recent observations of TRAPPIST-1e report stellar contamination affecting spectral features $>$ 3 $\mu$m, well past where this work models contamination and including regions where CO$_2$ and CH$_4$ features would be apparent \citep{Espinoza2025}. This further emphasizes the problem of stellar contamination and the need for further modeling work and new techniques to account for it. Work is being done to better account for stellar contamination in our observations and analyses.  \cite{Espinoza2025} was able to place constraints on the atmosphere of TRAPPIST-1e by implementing a multiplicative Gaussian Process (GP) to jointly model stellar contamination and atmospheric spectra while directly incorporating the limited knowledge of the star into the analysis. Other work has discussed using observations of TRAPPIST-1b as a proxy for stellar contamination, as 1b is most likely a bare rock and back-to-back observations could utilize the 1b transit to ``decontaminate" the data for the other planet \citep{Rathcke2025}. \textit{JWST} observations GO 6456 and 9256 will attempt this method with TRAPPIST-1e, and if successful, implementing similar methods for observations of 1f in the future could improve our abilities to constrain its atmosphere even without a complete understanding of the host star. Additionally, future retrieval codes may allow for stellar parameters to vary between visits while atmospheric parameters are held constant. With our current tools, a similar analysis would have been computationally impossible with \texttt{Multinest}, but future work may be able to better account for the star's changes between visits.

Stellar contamination makes the retrieval of rocky planet atmospheres at short wavelengths very difficult with current modeling techniques. While the most prominent features at longer wavelengths can still be retrieved or constrained for TRAPPIST-1f, it would take significant time and effort for \textit{JWST} to constrain the molecules key to gauging the presence of a potentially habitable atmosphere, as well as careful modeling of the stellar contamination. However, it is well within a small \textit{JWST} proposal to detect a majority-CO$_2$ atmosphere, which is likely a prerequisite for habitability and could represent a first step in a deeper study of TRAPPIST-1f's atmosphere. With 50 transits or more needed to begin to constrain features truly indicative of a habitable world, investigations into TRAPPIST-1f are still just beginning, and refinement of our modeling and observational techniques is needed as we continue to search the TRAPPIST-1 system for Earth-like conditions. 

\section*{Data Availability}

The atmospheric model used for TRAPPIST-1f can be found in \cite{Payne2024}. Our forward model spectra, simulated JWST observational data, and retrieval samples can be accessed at \url{https://doi.org/10.5281/zenodo.20836651}, and example notebooks for running \POSEIDON can be found at \url{https://poseidon-retrievals.readthedocs.io/en/latest/}.



\bibliographystyle{mnras}
\bibliography{main} 




\appendix

\section{Bayesian Evidence Table}

\begin{table*}
	\caption{Bayesian Evidence for atmospheric components in each retrieval variation.}
    \centering
	\label{tab:results}
    \begin{tabular}{lcccr}
        \textbf{Molecule} & \textbf{Model} & \textbf{Number of Transits} & \textbf{Bayes Factor} & \textbf{Detection Evidence} \\
        \hline
        \hline
        CO$_2$ & Uncontaminated & 5 + 5 & 5.72 & Weak \\
        CO$_2$ & Uncontaminated & 10 + 10 & 289 & Strong \\
        CO$_2$ & Uncontaminated & 15 + 15 & $2.38\times10^5$ & Detection \\
        \hline
        CO$_2$ & Contaminated & 5+ 5 & 6.26 & Weak \\
        CO$_2$ & Contaminated & 10 + 10 & 197 & Strong \\
        CO$_2$ & Contaminated & 15 + 15 & $1.04*10^4$ & Detection \\
        \hline
        \hline
        CH$_4$ & Uncontaminated & 5 + 5 & 0.902 & N/A \\
        CH$_4$ & Uncontaminated & 10 + 10 & 1.11 & N/A \\
        CH$_4$ & Uncontaminated & 15 + 15 & 1.83 & N/A \\
        CH$_4$ & Uncontaminated & 25 + 25 & 4.23 & Weak \\
        CH$_4$ & Uncontaminated & 50 + 50 & 120 & Moderate \\
        CH$_4$ & Uncontaminated & 100 + 100 & 1.09$\times$ 10$^5$ & Detection \\
        \hline
        CH$_4$ & Uncontaminated, NIRSpec Only & 5 & 0.878 & N/A \\
        CH$_4$ & Uncontaminated, NIRSpec Only & 10 & 1.10 & N/A \\
        CH$_4$ & Uncontaminated, NIRSpec Only & 15 & 1.73 & N/A \\
        CH$_4$ & Uncontaminated, NIRSpec Only & 25 & 4.02 & Weak \\
        CH$_4$ & Uncontaminated, NIRSpec Only & 50 & 100 & Moderate \\
        CH$_4$ & Uncontaminated, NIRSpec Only & 100 & 1.50$\times$ 10$^5$ & Detection \\
        \hline 
        CH$_4$ & Contaminated & 5 + 5 & 0.744 & N/A \\
        CH$_4$ & Contaminated & 10 + 10 & 0.883 & N/A \\
        CH$_4$ & Contaminated & 15 + 15 & 0.959 & N/A \\
        CH$_4$ & Contaminated & 25 + 25 & 1.17 & N/A \\
        CH$_4$ & Contaminated & 50 + 50 & 5.36 & Weak \\
        CH$_4$ & Contaminated & 100 + 100 & 177 & Strong \\
        \hline
        CH$_4$ & Contaminated, NIRSpec Only & 5 & 0.696 & N/A \\
        CH$_4$ & Contaminated, NIRSpec Only & 10 & 0.691 & N/A \\
        CH$_4$ & Contaminated, NIRSpec Only & 15 & 0.796 & N/A \\
        CH$_4$ & Contaminated, NIRSpec Only & 25 & 1.15 & N/A \\
        CH$_4$ & Contaminated, NIRSpec Only & 50 & 5.13 & Weak \\
        CH$_4$ & Contaminated, NIRSpec Only & 100 & 220 & Strong\\
        \hline
        \hline
        H$_2$O & Uncontaminated & 5 + 5 & 0.798 & N/A \\
        H$_2$O & Uncontaminated & 10 + 10 & 0.726 & N/A \\
        H$_2$O & Uncontaminated & 15 + 15 & 0.825 & N/A \\
        H$_2$O & Uncontaminated & 25 + 25 & 0.675 & N/A \\
        H$_2$O & Uncontaminated & 50 + 50 & 1.05 & N/A \\
        H$_2$O & Uncontaminated & 100 + 100 & 1.82 &  N/A \\
        \hline
        H$_2$O & Uncontaminated, NIRSpec Only & 5 & 0.743 & N/A \\
        H$_2$O & Uncontamintated, NIRSpec Only & 10 & 0.718 & N/A \\
        H$_2$O & Uncontaminated, NIRSpec Only & 15 & 0.647 & N/A \\
        H$_2$O & Uncontaminated, NIRSpec Only & 25 & 0.589 & N/A \\
        H$_2$O & Uncontaminated, NIRSpec Only & 50 & 0.492 & N/A \\
        H$_2$O & Uncontaminated, NIRSpec Only & 100 & 0.639 & N/A \\
        \hline
        H$_2$O & Contaminated & 5 + 5 & 0.666 & N/A \\
        H$_2$O & Contaminated & 10 + 10 & 0.758 & N/A \\
        H$_2$O & Contaminated & 15 + 15 & 0.976 & N/A \\
        H$_2$O & Contaminated & 25 + 25 & 0.407 & N/A \\
        H$_2$O & Contaminated & 50 + 50 & 0.550 & N/A \\
        H$_2$O & Contaminated & 100 + 100 & 0.913 & N/A \\
        \hline
        H$_2$O & Contaminated, NIRSpec Only & 5 & 0.781 & N/A \\
        H$_2$O & Contaminated, NIRSpec Only & 10 & 0.613 & N/A \\
        H$_2$O & Contaminated, NIRSpec Only & 15 & 0.617 & N/A \\
        H$_2$O & Contaminated, NIRSpec Only & 25 & 0.546 & N/A \\
        H$_2$O & Contaminated, NIRSpec Only & 50 & 0.472 & N/A \\
        H$_2$O & Contaminated, NIRSpec Only & 100 & 0.724 & N/A \\
        \hline
        O$_2$ & Uncontaminated & 100 + 100 & 1.03 & N/A \\
        O$_2$ & Contaminated & 100 + 100 & 0.642 & N/A \\
        \hline
        O$_3$ & Uncontaminated & 100 + 100 & 0.697 & N/A \\
        O$_3$ & Contaminated & 100 + 100 & 0.568 & N/A \\
    \end{tabular}
\end{table*}

\section{Stellar Contamination Parameter Retrieval}

\begin{figure*}
    \centering
    \includegraphics[width=\textwidth]{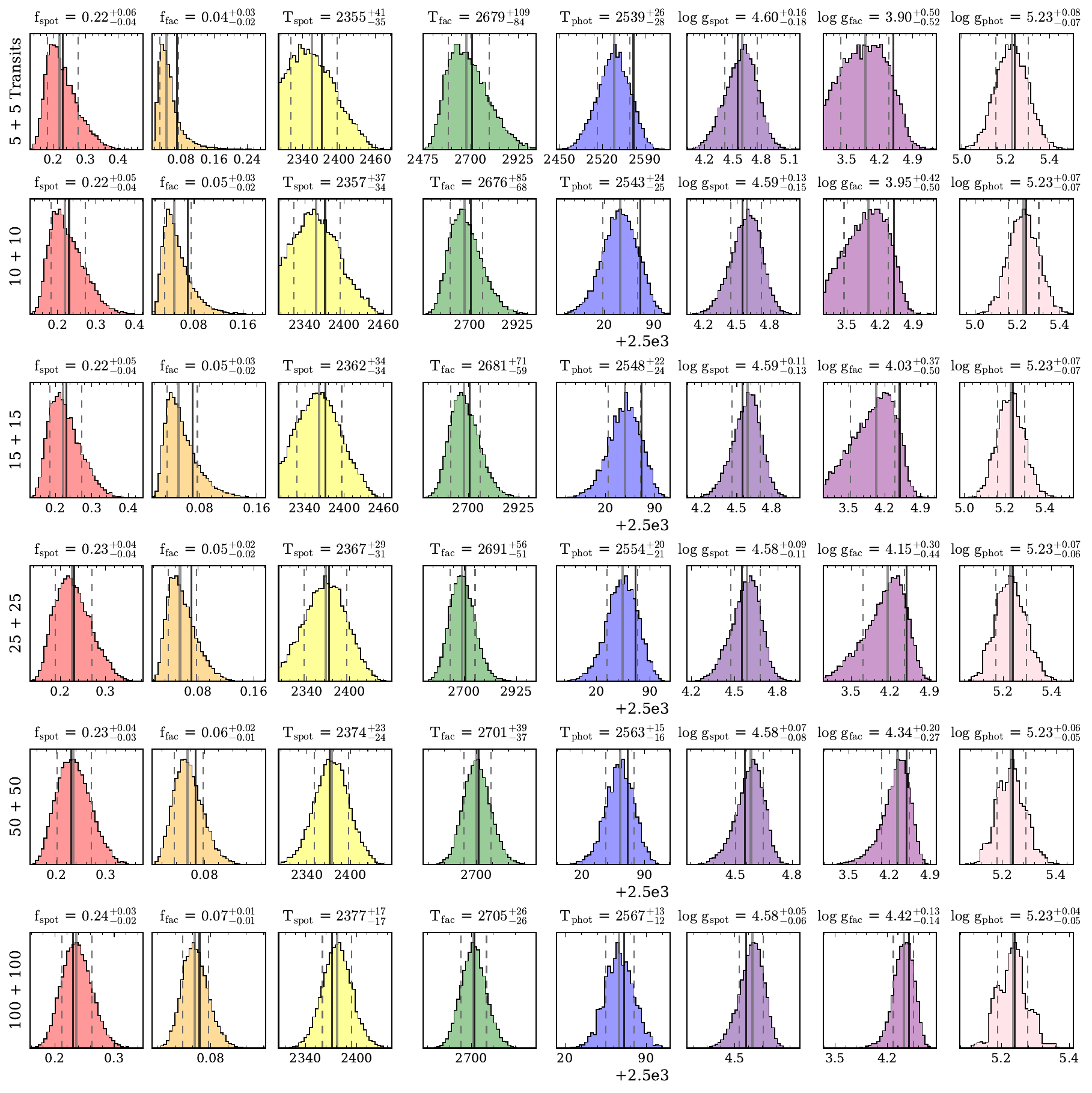}
    \caption{Retrieval results for stellar parameters in contaminated case.}
    \label{fig:stellar_param_retrieval}
\end{figure*}

\section{Effect of Gaussian Scatter on Retrievals}

\begin{figure*}
    \centering
    \includegraphics[width=\textwidth]{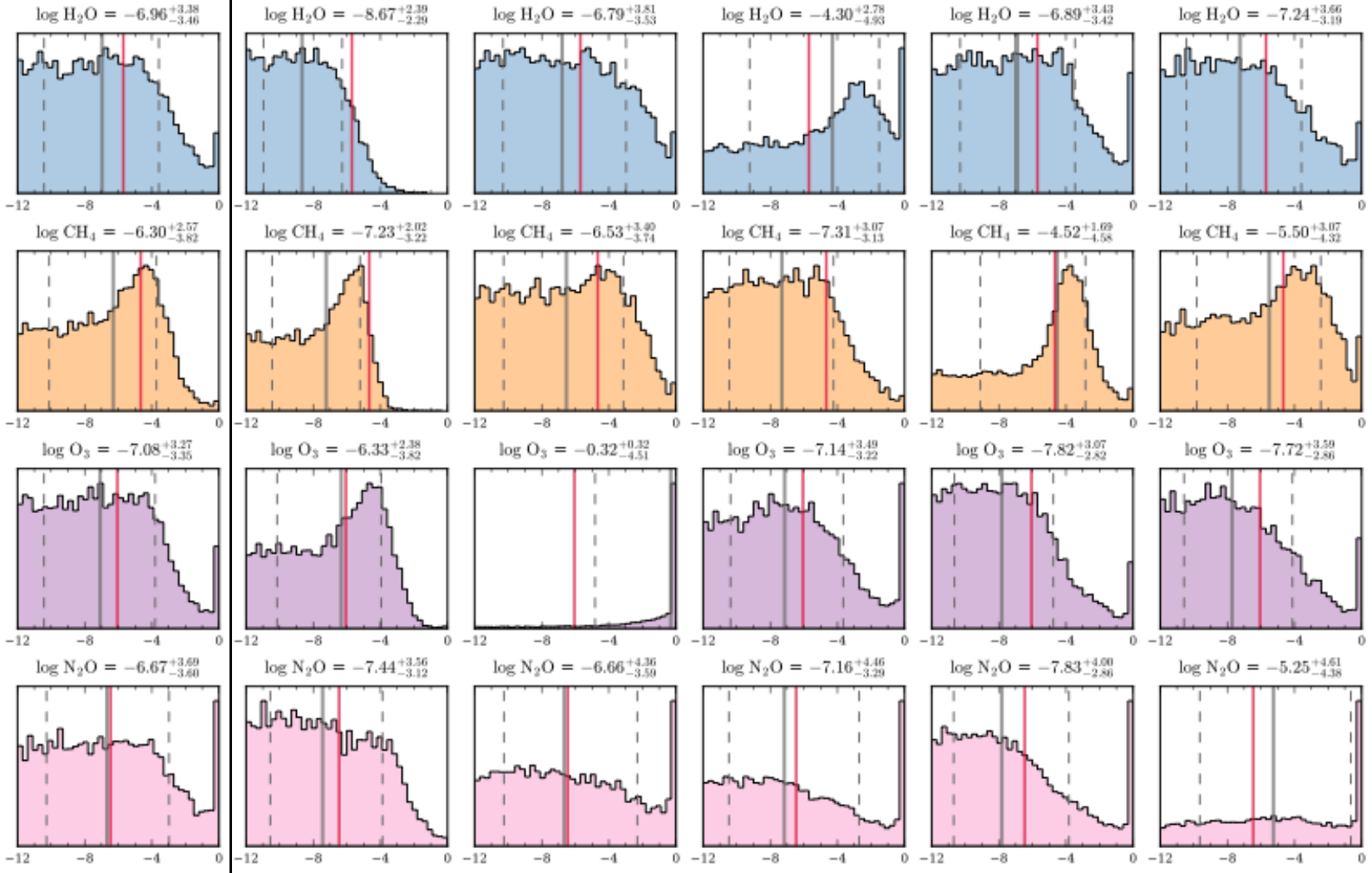}
    \caption{Retrieval results for the uncontaminated model, using 5+5 retrievals, of H$_2$O, CH$_4$, O$_3$, and N$_2$O. The first column, to the left of the black line, represents the unscattered results as shown in Figure \ref{fig:retrieved_histograms}, and the remaining five columns show retrieval results using data with differing Gaussian scatter draws. The result for O$_3$, in the third column, serves as a good example of how a draw of Gaussian noise can alter our results at a low number of transits.}
    \label{fig:appendix_uncontam_gaussian_scatter}
\end{figure*}

\begin{figure*}
    \centering
    \includegraphics[width=\textwidth]{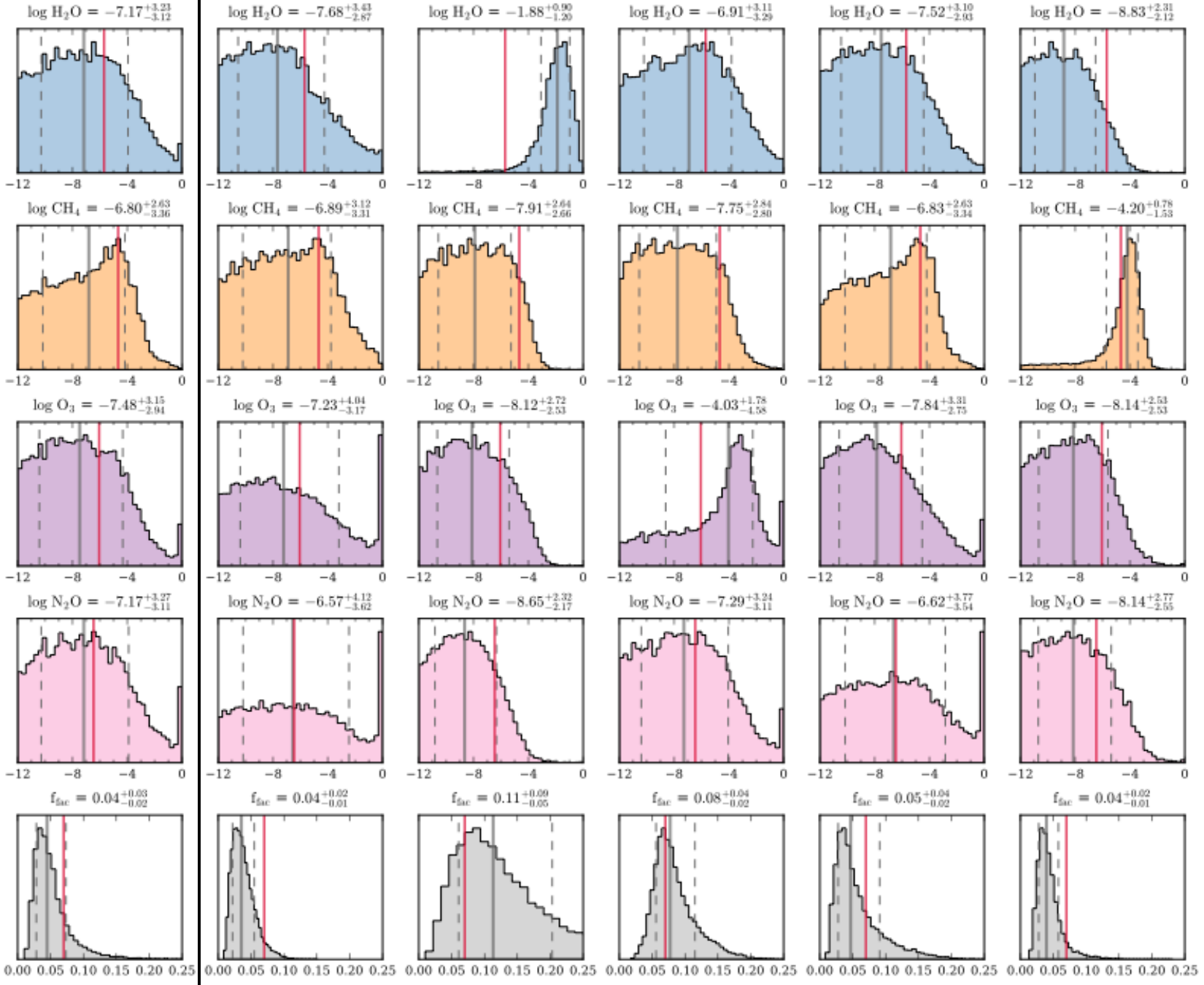}
    \caption{Retrieval results for the contaminated model, using 5+5 retrievals, of H$_2$O, CH$_4$, O$_3$, N$_2$O, and F$_{fac}$. The first column, to the left of the black line represents the unscattered results as shown in Figure \ref{fig:retrieved_histograms}, and the remaining five columns show retrieval results using data with differing Gaussian scatter draws.}
    \label{fig:appendix_contam_gaussian_scatter}
\end{figure*}


\bsp	
\label{lastpage}
\end{document}